\documentclass[aps]{revtex4-1}
\usepackage{amsmath}
\usepackage{latexsym}
\usepackage{float} 
\usepackage{amssymb} 
\usepackage{graphicx}
\usepackage{epsfig}
\begin{document} 
\title{Hydrophobic interactions with coarse-grained model for water}
\author{S. A. Egorov\footnote{Corresponding author e-mail: 
sae6z@virginia.edu}}
\affiliation{Department of Chemistry, University of 
Virginia, McCormick Road, Charlottesville, Virginia 22904, USA}
\begin{abstract}

Integral equation theory is applied to a coarse-grained model of water 
to study potential of mean force between hydrophobic solutes. 
Theory is shown to be in good agreement with
the available simulation data for methane-methane and 
fullerene-fullerene potential of mean force in water; the potential of
mean force is also decomposed into its entropic and enthalpic
contributions. Mode coupling theory is employed to compute
self-diffusion coefficient of water, as well as diffusion coefficient
of a dilute hydrophobic solute; good agreement with molecular dynamics
simulation results is found.

\end{abstract}
\maketitle
\newcommand{\nc}{\newcommand}
\newcommand{\tbf}{\textbf}
\nc{\cntr}[1]{\begin{center}{\bf #1}\end{center}}
\nc{\ul}{\underline} 
\nc{\fn}{\footnote}
\nc{\pref}{\protect\ref}
\setlength{\parindent}{.25in}
\nc{\stab}[1]{\begin{tabular}[c]{c}#1\end{tabular}}
\nc{\doeps}[3]{\stab{\setlength{\epsfxsize}{#1}\setlength{\epsfysize}{#2}
\epsffile{#3}}}
\nc{\sq}{^{2}}
\nc{\eps}{\epsilon}
\nc{\sig}{\sigma}
\nc{\lmb}{\lambda}
 \nc{\lbr}{\left[}
\nc{\rbr}{\right]}
\nc{\lpar}{\left(} 
\nc{\rpar}{\right)}
\nc{\ra}{\rightarrow}
\nc{\Ra}{\Rightarrow}
\nc{\lra}{\longrightarrow}
\nc{\LRa}{\Longrightarrow}
\nc{\la}{\leftarrow}
\nc{\La}{\Leftarrow}
\nc{\lla}{\longleftarrow}
\nc{\LLa}{\Longleftarrow}
\nc{\EE}[1]{\times 10^{#1}}
\nc{\simleq}{\stackrel{\displaystyle <}{\sim}}
\newcommand{\be}{\begin{equation}}
\newcommand{\ee}{\end{equation}}
\newcommand{\bea}{\begin{eqnarray}}
\newcommand{\eea}{\end{eqnarray}}
\newcommand{\R}{\vec{R}}
\newcommand{\bc}{\begin{center}}
\newcommand{\ec}{\end{center}}

\section{Introduction}
\label{sc1}

Hydrophobic interactions play an important role in a wide variety of
phenomena, including collapse of hydrophobic polymers in water, micelle
formation, and protein folding.\cite{berne09,pratt02} As a result, this
problem has been actively studied 
theoretically.\cite{pratt77,lee84,watanabe86,smith93,ludemann96,ludemann97,garde01,shimizu01,ghosh02,huang03,rajamani05,choudhury05,li05,li05b,howard08,chiu09,godawat09,athawale09,makowski10,setny10,chaimovich09,hammer10} 
The interactions between hydrophobic solutes in water have been
investigated using both computer simulations and mean field
theoretical methods. The former approach produces exact results for a
given microscopic model, but at a significant computational cost,
especially for large-size solutes, which require simulating a very
large number of water molecules. The latter approach is much less
computationally demanding, but inevitably involves approximations. 

One possible way to reduce computational cost associated with
simulation studies is to employ simplified coarse-grained solvent 
models.\cite{chiu09,chaimovich09,hammer10} Of particular relevance to
the present study is the recent work of Shell and 
co-workers,\cite{chaimovich09,hammer10} who have shown that several 
thermodynamic and dynamic anomalies of pure water, as well as various
features of hydrophobic interactions can be reasonably well reproduced
with a coarse-grained water model based on an isotropic pairwise
additive ``Lennard-Jones plus Gaussian'' (LJG) interaction potential. 
While the work of Shell and co-workers employed simulation-based
techniques, their microscopic model based on an 
isotropic pair potential is ideally suited for
approaches such as integral equation theory and mode-coupling theory
(MCT), which would reduce computational expenses even 
further.\cite{egorov08b,egorov08c,egorov08e}

Shell and co-workers have found that in order to properly reproduce 
the behavior of potential of mean force (PMF) between two hydrophobic
solutes in water (e.g. its variation with the solvent temperature),
the parameters of the coarse-grained LJG potential must be taken to be
dependent on the thermodynamic state point of the solvent. Nevertheless,
once the potential is parameterized for a given density and temperature
of water, it can be employed to study the hydrophobic interactions
between various solutes, and here is precisely where the computational
efficiency of the integral equation approach could be very helpful. 
However, as mentioned above, the theory is necessarily approximate
and its accuracy needs to be tested against the simulation data. 
The goal of the present study is to perform a comparison between
theory and simulation in obtaining structural and dynamical properties
of both pure water and dilute solutions of hydrophobic solutes. 

The remainder of the paper is organized as follows. 
In Section~\ref{sc2} we present our microscopic model 
and theoretical methods for calculating structural,
thermodynamic, and dynamical properties. The results are
presented in Section~\ref{sc3}. Section~\ref{sc4} concludes the paper. 

\section{Microscopic Model and Theory}
\label{sc2}

\subsection{Microscopic Model}
\label{sc21}

As a coarse-grained model for water, we 
consider a system comprised of spherical particles interacting via
an isotropic LJG pair potential 
of the following form:~\cite{chaimovich09,hammer10} 
\be
\phi_{s}(r)=4\epsilon\left[\left(\frac{\sigma}{r}\right)^{12}
-\left(\frac{\sigma}{r}\right)^{6}\right]
+B\exp\left[-\frac{(r-r_0)^2}{\Delta^{2}}\right],
\label{phis}
\ee
where $\epsilon$ and $\sigma$ are the usual LJ parameters setting the
energy and length scale of the interaction, while parameter $B$ sets
the strength of the Gaussian term, whose center and width are
determined by $r_0$ and $\Delta$, respectively. 

In studying hydrophobic interactions, we consider dilute solutions,
where solutes interact with solvent particles via isotropic pair
potential $\phi(r)$ and with each other via another isotropic pair 
potential $\phi_{uu}(r)$,
particular forms of these potentials depend on the solute and
will be specified for each solute in Section~\ref{sc3}.

\subsection{Structural Properties}
\label{sc22}

We employ integral equation theory~\cite{hansen86} to compute 
the solvent-solvent ($g_s(r)$) and 
the solute-solvent ($g(r)$) pair distribution functions and 
the solute-solute PMF ($W(r)$).  

The starting point of the calculation is the
Ornstein-Zernike relation between the solvent-solvent 
total ($h_s(r)=g_s(r)-1$) and direct ($c_s(r)$) pair correlation
functions:~\cite{hansen86}
\be
h_s(r)=c_s(r)+\rho\int d\vec{r}^{\,\prime}h_s(r^{\,\prime})
c_s(|\vec{r} - \vec{r}^{\,\prime}|),
\label{oz}
\ee
where $\rho$ is the bulk solvent number density. In order to solve the
above equation, one needs an additional closure relation between 
$h_s(r)$ and $c_s(r)$. Here we employ thermodynamically
self-consistent hybrid mean spherical approximation (HMSA) 
closure:~\cite{zerah86} 
\be
c_s(r)=\exp\{-\beta \phi_{s}(r)\} 
\left[
1+\frac{\exp\left[f(r)\left(h_s(r)-c_s(r)\right)\right]-1}{f(r)}\right]
-h_s(r_{1})+c_s(r_{1})-1,
\label{hmsa}
\ee
where $\beta=1/k_BT$, and $f(r)=1-\exp(-\alpha r)$. 
Note that this closure interpolates between the
soft-core mean spherical 
approximation\cite{madden80} ($\alpha \rightarrow 0$) and the 
(hypernetted chain) HNC closure  
($\alpha \rightarrow \infty$).\cite{zerah86}

The value of the parameter $\alpha$ is determined by a 
thermodynamic consistency condition that  is   based on equating the 
isothermal compressibilities, $\chi_T^v$ and $\chi_T^c$, 
that follow respectively from the virial 
and compressibility routes:~\cite{zerah86}
\bea
(\rho k T \chi_{T}^{v})^{-1}&=&
\left(\frac{\partial \beta P^{v}}{\partial \rho}\right), \nonumber \\
(\rho k T \chi_{T}^{c})^{-1}&=&
1-\rho \int d\vec{r} c_s(r),
\label{cons}
\eea
where the virial pressure $P^{v}$ is given by 
\be
\beta P^{v}=\rho - \frac{\beta \rho^2}{6} \int d\vec{r}r 
\phi^{\prime}_{s}(r)g_s(r),
\label{press}
\ee
with prime denoting differentiation with respect to the argument. 

For an infinitely dilute solute in a solvent, 
the solute-solvent total ($h(r)=g(r)-1$) and direct 
($c(r)$) pair correlation functions are also related via
Ornstein-Zernike equation:
\be
h(r)=c(r)+\rho\int d\vec{r}^{\,\prime}h(r^{\,\prime})
c_s(|\vec{r} - \vec{r}^{\,\prime}|),
\label{oz1}
\ee
We employ the HNC closure to solve for $h(r)$ and $c(r)$:
\be
h(r)=\exp[-\beta\phi(r)+h(r)-c(r)]
\label{hnc}
\ee
The accuracy of this approach will be tested in the next Section by
comparing theoretical results with simulations.

Finally, for two dilute solutes in a solvent, the solute-solute PMF is
composed of two terms:
\be
W(R)=\phi_{uu}(R)+\Delta W(R),
\label{pmf}
\ee
where the first term is the bare solute-solute interaction potential, 
while the second term is the solvent-mediated PMF given by the following exact
relations:~\cite{egorov01c} 
\begin{equation}
\Delta W(R)=\int_{R}^{\infty}F(R^{\,\prime})dR^{\,\prime},
\label{wr}
\end{equation}
where the excess mean force, $F(R)$, is given by:
\begin{equation}
\vec{F}(R)= -\int d\vec{r} {{~}} \nabla \phi(r)
\rho(\vec{r};R).
\label{fr}
\end{equation}
In the above, $\rho(\vec{r};R)$ is the conditional
probability of finding the solvent particle at $\vec{r}$ given that
one solute is at the origin, and the other solute is located at
$\vec{R}$. We compute this conditional probability from the
anisotropic HNC 
closure, which has been thoroughly tested against computer simulations
in our earlier work:~\cite{egorov01c,egorov02e} 
\be
\rho(\vec{r};R) =  \rho \exp\left[-\beta
\left(\phi(r)+\phi(|\vec{r}-\vec{R}|)\right)+
 \int d \vec{r}^{\,\prime} c(|\vec{r} - \vec{r}^{\,\prime}|)
(\rho(\vec{r}^{\,\prime};R) - \rho) \right].
\label{rho}
\ee
From the solute-solute PMF one can easily obtain the solute-solute
radial distribution function given by:
\be
g_{uu}(R)=\exp(-\beta W(R)).
\label{guur}
\ee

As will be seen from the results presented below, for a typical
hydrophobic solute such as methane, the solute-solute PMF typically
displays two pronounced minima corresponding to the contact pair and
the solvent-separated pair of the solutes.\cite{ludemann96,ghosh02} 
Concomitantly, the solute-solute radial distribution function is
characterized by two corresponding maxima, with the second maximum 
(corresponding to the solvent-separated pair) bracketed by two minima
located at $R_1$ and $R_2$.
In what follows, we will study the dependence of the methane-methane PMF
and $g_{uu}(R)$ on the water density and temperature. 
In order to characterize the behavior of these functions 
by a single parameter, we will compute the
equilibrium constant for the conversion of the contact pair into the 
solvent-separated pair:\cite{ludemann96} 
\begin{equation}
K_{eq}=\frac{\int_{0}^{R_{1}}g_{uu}(r)r^{2}dr}
{\int_{R_{1}}^{R_{2}}g_{uu}(r)r^{2}dr}.
\label{keq}
\end{equation}

\subsection{Dynamical Properties}
\label{sc23}

We employ the MCT-based approach to compute the
self-diffusion coefficient of the pure water, as well as the
diffusion coefficient of a dilute solute. 
In this approach, 
the self-diffusion coefficient is obtained from the total 
friction $\zeta$:\cite{balucani94,bhattacharyya98,egorov03b}   
\be
D=\frac{k_BT}{m_s\zeta},
\label{dc}
\ee
where $m_{s}$ is the mass of the water molecule, and the MCT
expression for $\zeta$ is comprised of binary and 
collective density contributions:\cite{bhattacharyya97}   
\be
\zeta=\zeta_b+\zeta_{\rho\rho}
\label{zeta} 
\ee
In what follows, we will be calculating self-diffusion coefficient at
relatively high water densities, and therefore we neglect 
the hydrodynamic contribution to friction which 
arises from the coupling of the tagged solvent particle  
motion to the collective transverse current mode.\cite{bhattacharyya97}    

The binary term is given by the total time integral of the fast
decaying binary component of the time dependent friction, which  
we model as a Gaussian function, whose parameters
are chosen to reproduce the exact short-time behavior of the total
time-dependent friction, $\zeta(t)$, 
up to the term of order $t^2$:\cite{balucani94}   
\be
\zeta_{b}(t)=\zeta(0)\exp\left[\frac{\ddot{\zeta}(0)t^2}
{2\zeta(0)}\right],
\label{zetab} 
\ee
with the initial-time value given by:  
\be
\zeta(0)=\frac{4\pi\rho}{3m_s}\int_{0}^{\infty}dr r^2 g_s(r)
\nabla^2\phi_s(r).
\label{zetab0}
\ee
The zero-time value of the second time derivative, $\ddot{\zeta}(0)$, 
can be similarly obtained from the fluid
intermolecular potential and radial distribution 
function.\cite{egorov03b,balucani94}   

The density term arises 
from the coupling of the tagged solvent particle motion 
to the collective density mode:\cite{bhattacharyya97}    
\be
\zeta_{\rho\rho}=\int_{0}^{\infty}\zeta_{\rho\rho}(t)dt,
\label{zetarr}
\ee
with
\be
\zeta_{\rho\rho}(t)=\frac{k_BT\rho}{6\pi^2m_s}
\int_{0}^{\infty}dk k^4 c_s(k)^2 
\left[F_s(k,t)F(k,t)-F_{s}^{0}(k,t)F^{0}(k,t)\right]
\label{zetarrt}
\ee
where $F(k,t)$ is the solvent dynamic structure factor, 
$F_{s}^{0}(k,t)=\exp(-k_BTk^2t^2/2m_s)$, and 
$F_s(k,t)$ is the solvent self-dynamic structure 
factor, for which we have adopted a simple Gaussian 
model:\cite{balucani94,bhattacharyya97}    
\be
F_s(k,t)=\exp\left[\frac{-k_BTk^2}{m_s\zeta}
\left(t+\frac{1}{\zeta}(e^{-t\zeta}-1)\right)\right].
\label{fskt}
\ee 

We obtain the solvent dynamic structure factor from the 
continued fraction representation of its Laplace 
transform truncated at the second order:\cite{balucani94}    
\be
F(k,z)=
\frac{S(k)}
{z+\frac{\displaystyle{\delta_1(k)}}
{\displaystyle{z+}\frac{\displaystyle{\delta_2(k)}}
{\displaystyle{z+}\displaystyle{\tau^{-1}(k)}},
}}
\label{fkz}
\ee
where $\delta_i(k)$ is the initial time value of the $i^{\mbox{th}}$
order memory function (MF) of $F(k,t)$.
For the parameter 
$\tau^{-1}(k)$ we use the expression due to Lovesey:\cite{lovesey71}  
$\tau^{-1}(k)=2\sqrt{\delta_2(k)/\pi}$. 
The quantities $\delta_1(k)$ and $\delta_2(k)$  can be easily calculated
from the first three short-time expansion coefficients of $F(k,t)$; the
microscopic expressions for the latter are well-known and will not be
reproduced here.\cite{hansen86,balucani94,bansal77}

Given that the self-dynamic structure factor in Eq.~(\ref{fskt})  
is a function of $\zeta$,
which, in turn, depends on $F_s(k,t)$ via 
Eq.~(\ref{zetarr}) and (\ref{zetarrt}), the 
above set of MCT equations for $\zeta$ needs 
to be solved iteratively and
self-consistently. One could use a more accurate model for $F_s(k,t)$
in terms of the velocity time correlation function of a tagged fluid
particle,\cite{bhattacharyya98} but our numerical calculations have
shown that this does not change the results for $D$ in a noticeable
way.     

The diffusion coefficient of a dilute solute is obtained in the same
way as the solvent self-diffusion coefficient, with the understanding
that in the above set of MCT equations $\phi_s(r)$, $g_s(r)$, and 
$c_s(r)$ are replaced with $\phi(r)$, $g(r)$, and $c(r)$,
respectively. Additionally, 
$F_s(k,t)$ now has the meaning of the solute self-dynamic structure 
factor, and 
$F_{s}^{0}(k,t)=\exp(-k_BTk^2t^2/2m)$ is its inertial component, where
$m$ is the solute mass. Finally, $m_s$ is replaced with $m$ in 
Eqs.~(\ref{dc}), (\ref{zetab0}), (\ref{zetarrt}), and (\ref{fskt}).

In the next section, we compare our theoretical results 
for structural and dynamical properties of pure water 
and dilute solutions of hydrophobic solutes with MD data and analyze the 
dependence of various computed properties on the solvent thermodynamic
conditions.
 
\section{Results}
\label{sc3}

\subsection{Pure Water}
\label{sc31}

We start by presenting our results for the structural and dynamical
properties of pure water. Parameters entering the LJG potential given
by Eq.~(\ref{phis}) are dependent on density and temperature and are
taken from Ref.~\onlinecite{chaimovich09}, where they were obtained
from the relative entropy optimization method.  
In Fig.~1 we compare simulation results
  for the oxygen-oxygen radial distribution function of all-atom SPC/E 
water,\cite{athawale09} and the HMSA result for $g_s(r)$ 
of coarse-grained water described by the LJG potential.
The two distribution functions are generally in good agreement except
some discrepancies in the range between 3 and 6 \AA. The same
discrepancies were seen in the earlier comparison of simulated 
$g_s(r)$ for SPC and LJG water, i.e. if one were to compare HMSA
theory with simulations done for the coarse-grained LJG water (not
shown), the two distribution functions would be nearly
indistinguishable. 

\begin{figure}[htb]
\includegraphics[scale=0.5]{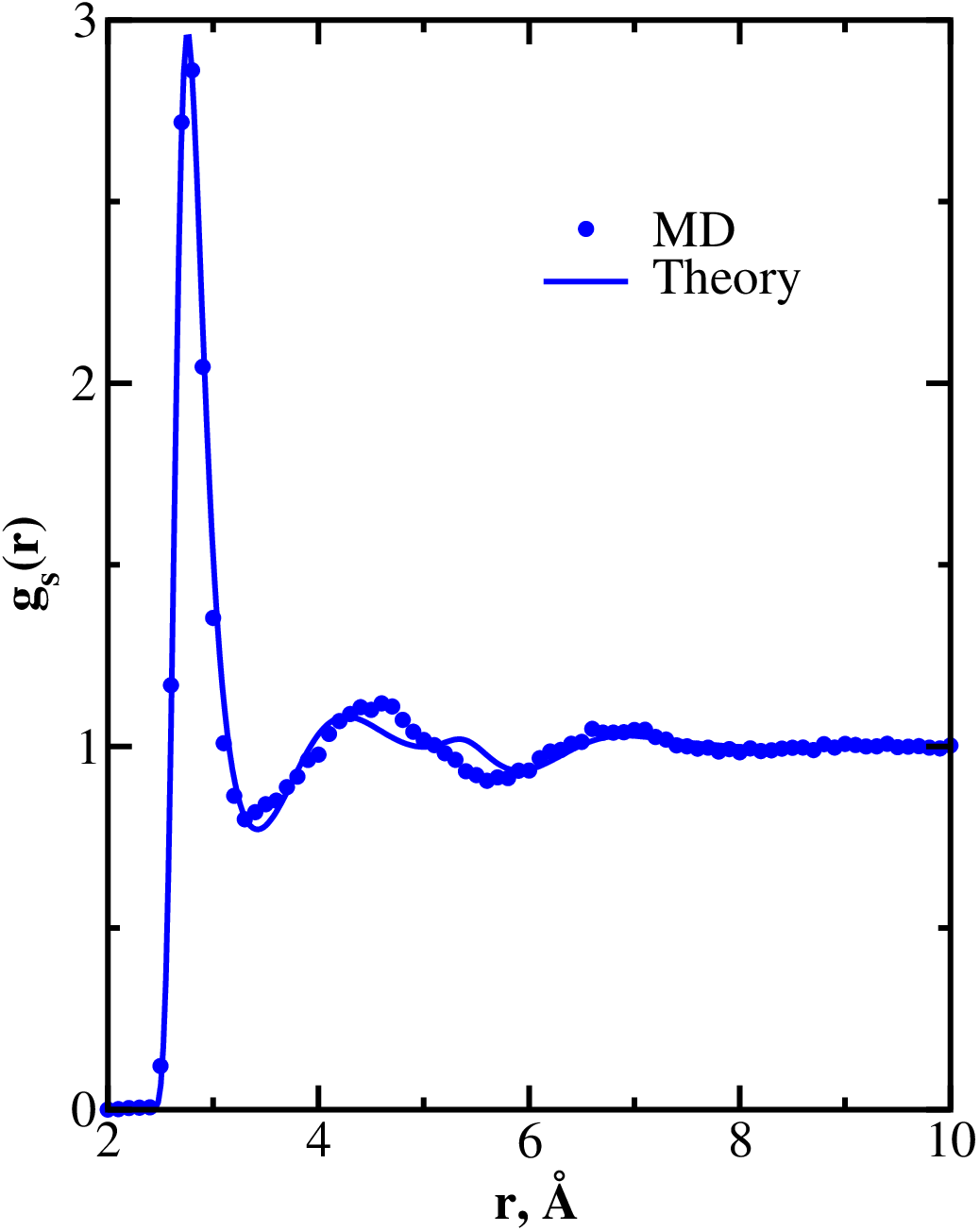}
\vspace{0.5cm}
\caption{Radial distribution function of pure water at T=300~K and 
$\rho$=1 g/cm$^3$. Symbols are the simulation results
  for the oxygen-oxygen radial distribution function of all-atom SPC/E 
water,\cite{athawale09} and the line is from HMSA theory for
coarse-grained water described by the LJG potential.}
\label{fig1}
\end{figure}

The same level of agreement between theoretical and
simulation results for the structural properties of coarse-grained
water is obtained at other thermodynamic
conditions, as can be seen from Fig.~2a, where we 
present MD\cite{chaimovich09} (symbols)
and HMSA (lines) pressure isochores for two values of water density: 
$\rho$=0.9 g/cm$^3$ and 1 g/cm$^3$, with theoretical
results obtained from Eq.~(\ref{press}). 
\begin{figure}
\includegraphics[scale=0.43]{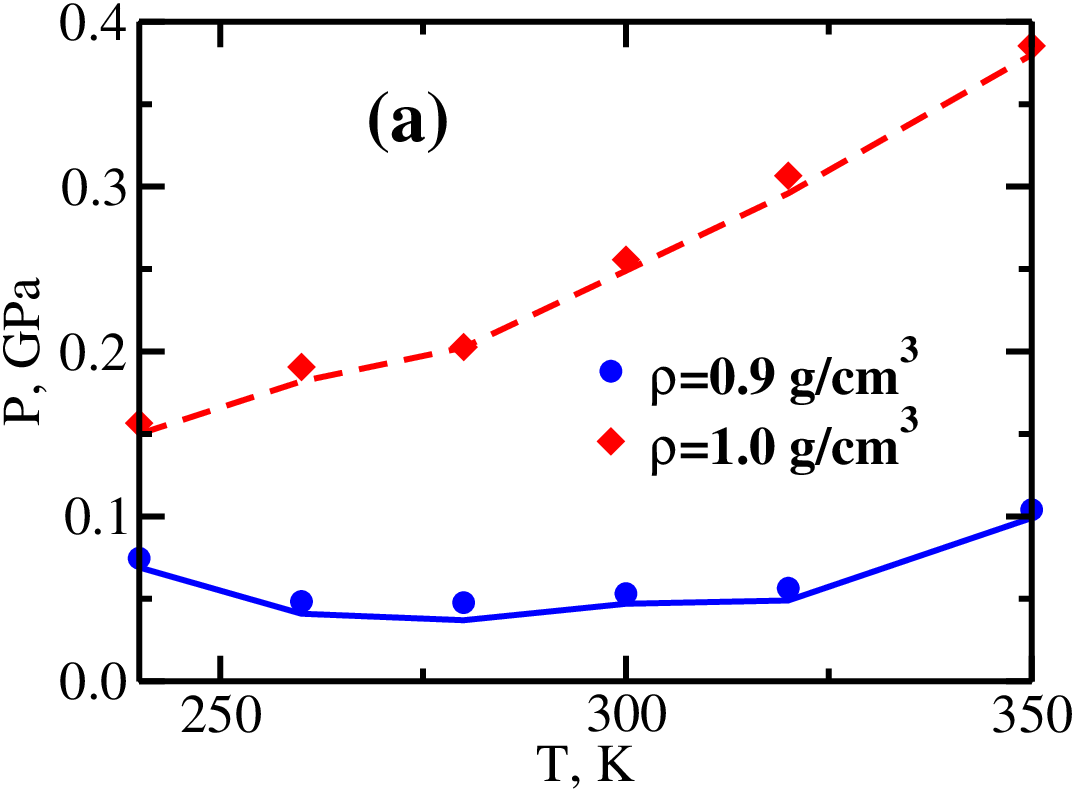}
\hspace{0.50cm}
\includegraphics[scale=0.43]{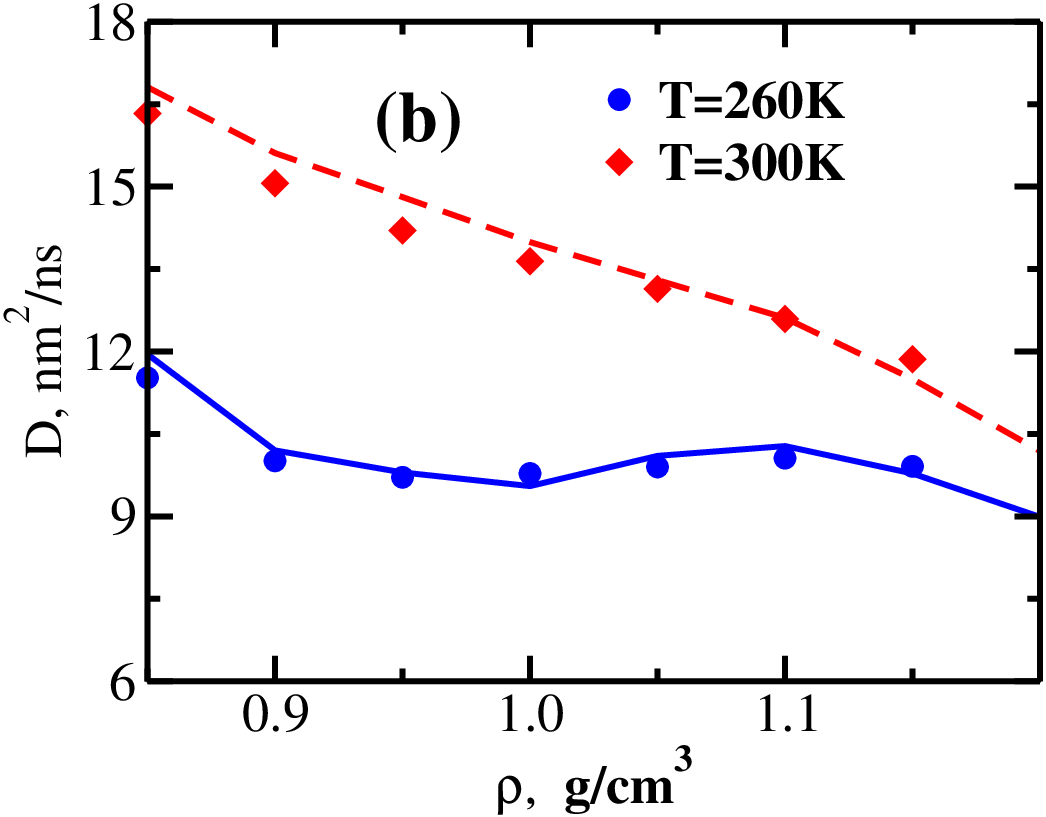}
\caption{(a)Pressure isochores for the coarse-grained LJG water model;
 symbols are from MD simulations,\cite{chaimovich09} and lines are
 from the HMSA integral equation theory.
(b) Self-diffusion coefficient of the coarse-grained LJG water as a
 function of density along two isotherms; symbols 
 are from MD simulations,\cite{chaimovich09} and lines are
 from the MCT approach. } 
\label{fig2}
\end{figure}
One sees that the HMSA results for pressure are in excellent
agreement with the MD data for both isochores 
throughout the entire temperature range studied, thereby confirming
the accuracy of the theory in providing structural information for the
coarse-grained water model. In particular, theory successfully
reproduces the characteristic anomalous water behavior, with pressure
decreasing as a function of $T$ along the isochore $\rho$=0.9 g/cm$^3$  
at the lower end of the temperature range studied, subsequently
passing through a minimum, and then growing with temperature. 

Having ascertained the accuracy of the HMSA theory  
in calculating structural
and thermodynamic properties of pure water, we
next use the structural data as input to compute water self-diffusion 
coefficient from the MCT approach. 
In Fig.~2b we present simulation\cite{chaimovich09} (symbols) 
and theoretical (lines) results for 
self-diffusion coefficient of the coarse-grained LJG water as a
function of density along two isotherms: T=260~K and T=300~K. 
At the higher temperature, the self-diffusion coefficient
decreases monotonically with increasing density, while at the lower
temperature, anomalous behavior is observed,
where the self-diffusion coefficient passes through a minimum around 
$\rho$=1.0 g/cm$^3$, increases with density between 1.0 and 1.1 g/cm$^3$,
passes through a maximum, and then decreases with density.
One sees that theory is in good agreement with simulation for both
isotherms, and in particular captures the characteristic water 
diffusion anomaly well. 

Having considered structural, thermodynamic, and dynamic properties of
pure coarse-grained LJG water, we next turn to dilute solutions of
hydrophobic solutes in LJG water.  

\subsection{Methane Solutes in Water}
\label{sc32}

For the purpose of modeling dilute solution of methane in water, we
have employed LJ potentials for both solute-solvent and solute-solute
interactions:
\be
\phi(r)=4\epsilon_{us}\left[\left(\frac{\sigma_{us}}{r}\right)^{12}
-\left(\frac{\sigma_{us}}{r}\right)^{6}\right],
\label{phimeth}
\ee
and
\be
\phi_{uu}(r)=4\epsilon_{uu}\left[\left(\frac{\sigma_{uu}}{r}\right)^{12}
-\left(\frac{\sigma_{uu}}{r}\right)^{6}\right].
\label{phiuumeth}
\ee
In order to compare our theoretical results with the earlier
simulation study of methane solution in LJG 
water,\cite{chaimovich09,athawale09} 
we have chosen the same values of $\epsilon$ and $\sigma$ parameters
as in the simulations: 
$\sigma_{us}$=3.45~\AA, $\epsilon_{us}$=0.89~kJ/mol,   
$\sigma_{uu}$=3.73~\AA, and $\epsilon_{uu}$=1.23~kJ/mol.

\begin{figure}[htb]
\includegraphics[scale=0.5]{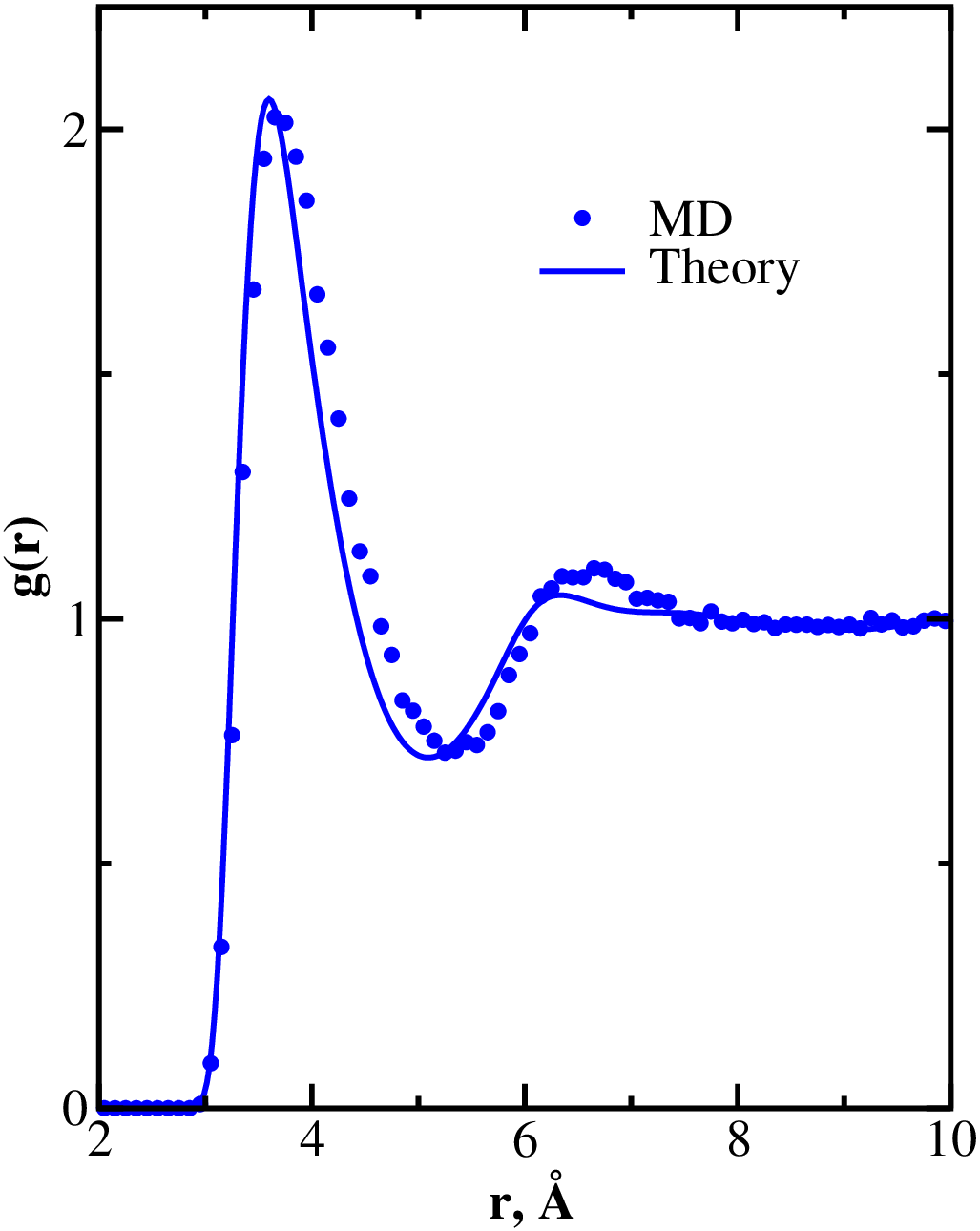}
\vspace{0.5cm}
\caption{Methane-water radial distribution function at T=300~K and 
$\rho$=1 g/cm$^3$. Symbols are the simulation results
  for the methane-water radial distribution function with all-atom SPC/E 
water,\cite{athawale09} and the line is from HNC theory for
coarse-grained water described by the LJG potential.}
\label{fig3}
\end{figure}
In Fig.~3 we compare simulation results
  for the methane-water radial distribution function with all-atom SPC/E 
water,\cite{athawale09} and the HNC result for $g(r)$ 
with coarse-grained LJG water.
The two distribution functions are generally in good agreement except
some discrepancies in the range between 4 and 7 \AA. 
As in the case of pure water, the agreement would be significantly
better if one were to compare theory with simulation performed for
methane in coarse-grained LJG water. 
\begin{figure}
\includegraphics[scale=0.43]{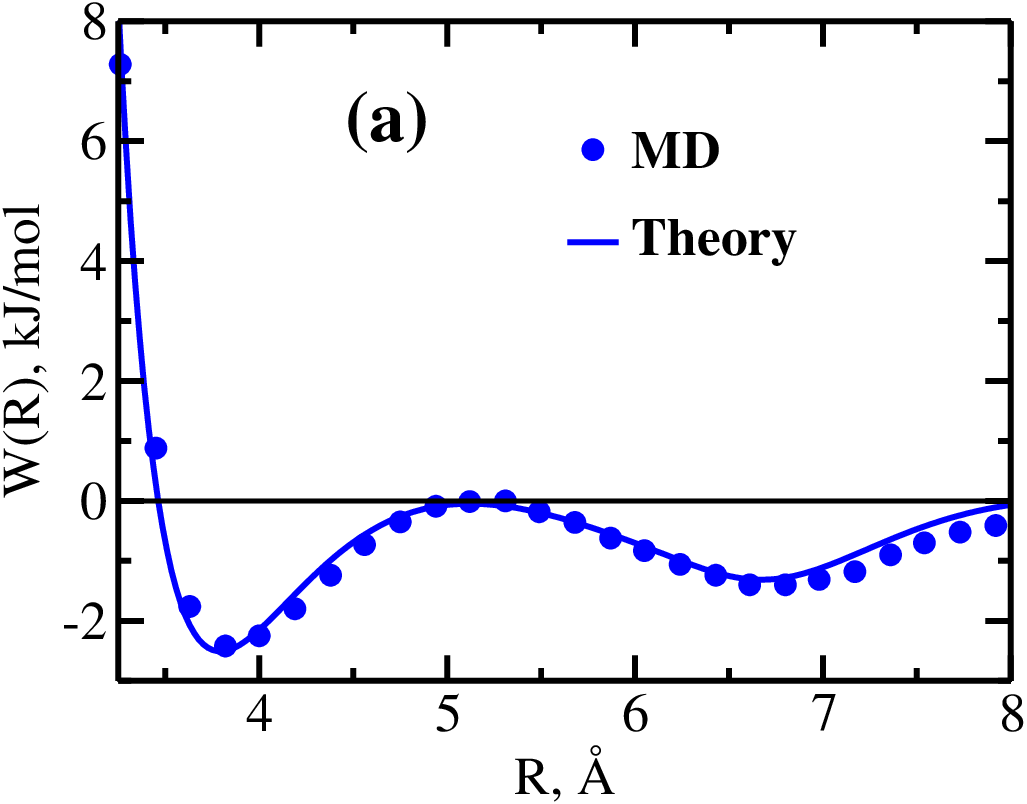}
\hspace{0.50cm}
\includegraphics[scale=0.43]{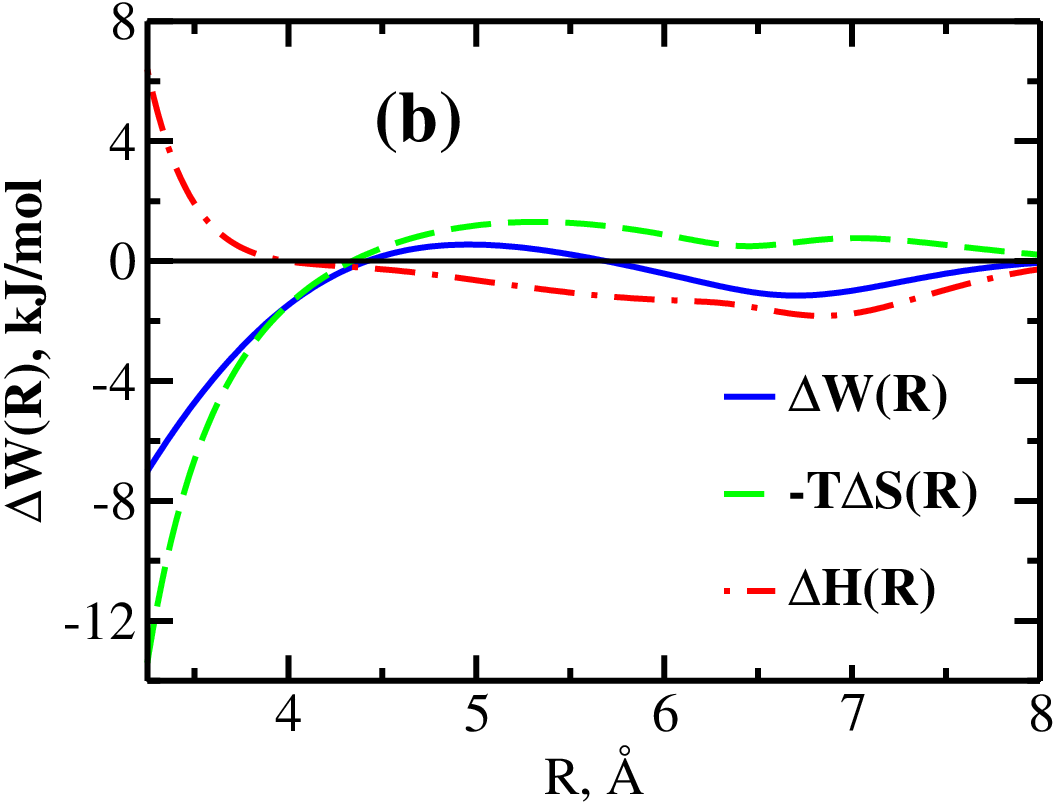}
\caption{(a)PMF between two dilute methane solutes 
 in coarse-grained LJG water;
 symbols are from MD simulations,\cite{chaimovich09} and lines are
 from the integral equation theory.
(b) Solvent-mediated PMF between two dilute methane solutes 
 in coarse-grained LJG water; also shown are entropic and enthalpic terms.} 
\label{fig4}
\end{figure}
This is confirmed by comparing
theoretical and simulation results for the PMF
between two dilute methane solutes in LJG water, which are presented
in Fig.~4a for thermodynamic state point T=300~K and $\rho$=1 g/cm$^3$.
One sees that theory is in excellent agreement with
simulation throughout the entire range of separations between the two
methane solutes, indicating that it accurately reproduces the
anisotropic density distribution of LJG water molecules around two
methane solutes at all separations considered. In particular, theory
properly reproduces the locations and the depths of the two minima in
the PMF -- the first one corresponding to the contact pair 
(located around 3.9 \AA), and the second one corresponding to the 
solvent-separated pair (located around 6.7 \AA); the contact pair
minimum is deeper compared to the solvent-separated one. 

The agreement is
equally good at the other two temperatures for which simulations were
performed: T=280~K and T=320~K (not shown). On the basis of this
information, we have decomposed the solvent-mediated PMF 
given by the integral equation theory into
enthalpic and entropic contributions:\cite{smith93,ghosh02}
\be
\Delta S(T)\approx\frac{\Delta W(T+\Delta T)-\Delta W(T-\Delta T)}
{2\Delta T},
\label{deltas}
\ee 
 and
\be
\Delta H=\Delta W + T\Delta S,
\label{deltah}
\ee 
where $\Delta T$=20~K in our calculations. The corresponding results
are presented in Fig.~4b. 
In agreement with previous simulation results,\cite{ghosh02}
one sees that the solvent-mediated PMF at
the location of the contact pair minimum is completely dominated by
the entropic contribution, with enthalpic term very close to zero but
slightly unfavorable (positive). At larger separations, the enthalpic
contribution becomes favorable and the entropic term unfavorable; 
the solvent-mediated PMF at
the location of the solvent-separated pair minimum is dominated by
the enthalpic contribution.\cite{ghosh02}

In order to present our results for methane-methane PMF at other
densities and temperatures in a compact form, we have computed the 
equilibrium constant for the conversion of the contact pair into the 
solvent-separated pair from Eq.~(\ref{keq}). Theoretical results for
$K_{eq}$ along two isotherms are presented in Fig.~5 together with the
only simulation data point available\cite{chaimovich09} at 
T=300~K and $\rho$=1 g/cm$^3$.
\begin{figure}[htb]
\includegraphics[scale=0.5]{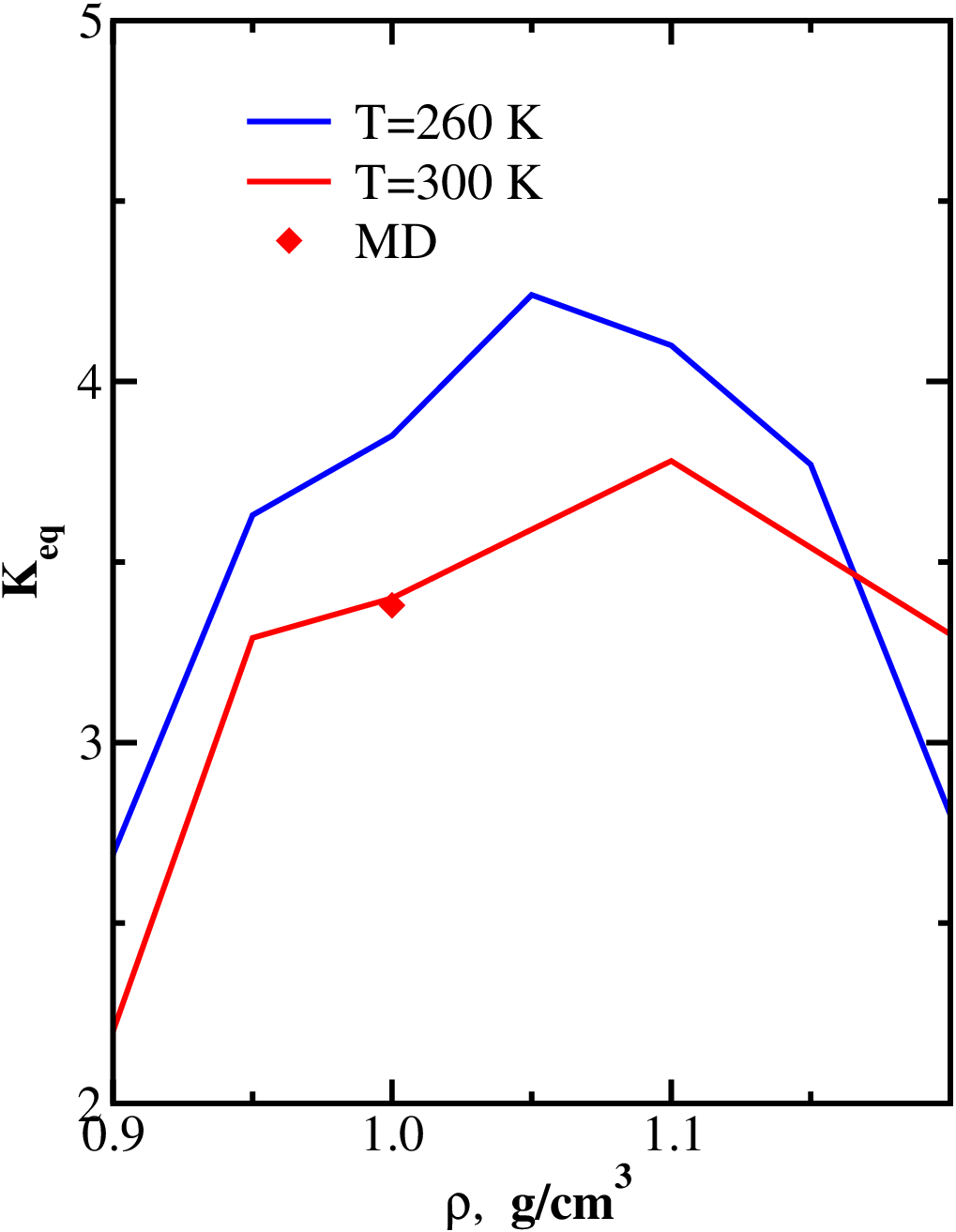}
\vspace{0.5cm}
\caption{Theoretical results for the 
equilibrium constant for the conversion of the contact pair into the 
solvent-separated pair along two isotherms.}
\label{fig5}
\end{figure}
One sees that in agreement with earlier simulation 
work,\cite{ludemann96} the solvent-separated pair is more stable in
terms of $K_{eq}$ at all densities and temperatures studied. For both
isotherms, the equilibrium constant first increases with density,
passes through a maximum between 1.05 g/cm$^3$ and 1.1 g/cm$^3$, and
then decreases with density. Except for the highest density
considered ($\rho$=1.2 g/cm$^3$), the equilibrium constant is larger
at the lower temperature. 

\subsection{Fullerenes in Water}
\label{sc33}

For the purpose of modeling dilute solution of fullerenes in water, we
have employed the following form for the solute-solvent interaction 
potential:\cite{girifalco00}
\be
\phi(r)=4 N \epsilon_{us}\frac{\sigma_{us}^{2}}{rR}
\left(\frac{1}{20}
\left[\left(\frac{\sigma_{us}}{R-r}\right)^{10}
-\left(\frac{\sigma_{us}}{R+r}\right)^{10}\right]
-\frac{1}{8}
\left[\left(\frac{\sigma_{us}}{R-r}\right)^{10}
-\left(\frac{\sigma_{us}}{R+r}\right)^{10}\right]
\right),
\label{phic60}
\ee
where $N$=60, R=3.55 \AA, $\sigma_{us}$=3.19 \AA, and 
$\epsilon_{us}$=0.392 kJ/mol.\cite{li05b}

The fullerene-fullerene direct interaction potential is given 
by:\cite{girifalco00}
\be
\phi_{uu}(r)=-\alpha
\left[\frac{1}{s(s-1)^3}+\frac{1}{s(s+1)^3}-\frac{2}{s^4}\right]
+\beta\left[\frac{1}{s(s-1)^9}+\frac{1}{s(s+1)^9}-\frac{2}{s^{10}}\right],
\label{phiuuc60}
\ee
where $\alpha$=4.4775~kJ/mol, $\beta$=0.0081~kJ/mol, and $s=r/2R$. 

\begin{figure}[htb]
\includegraphics[scale=0.5]{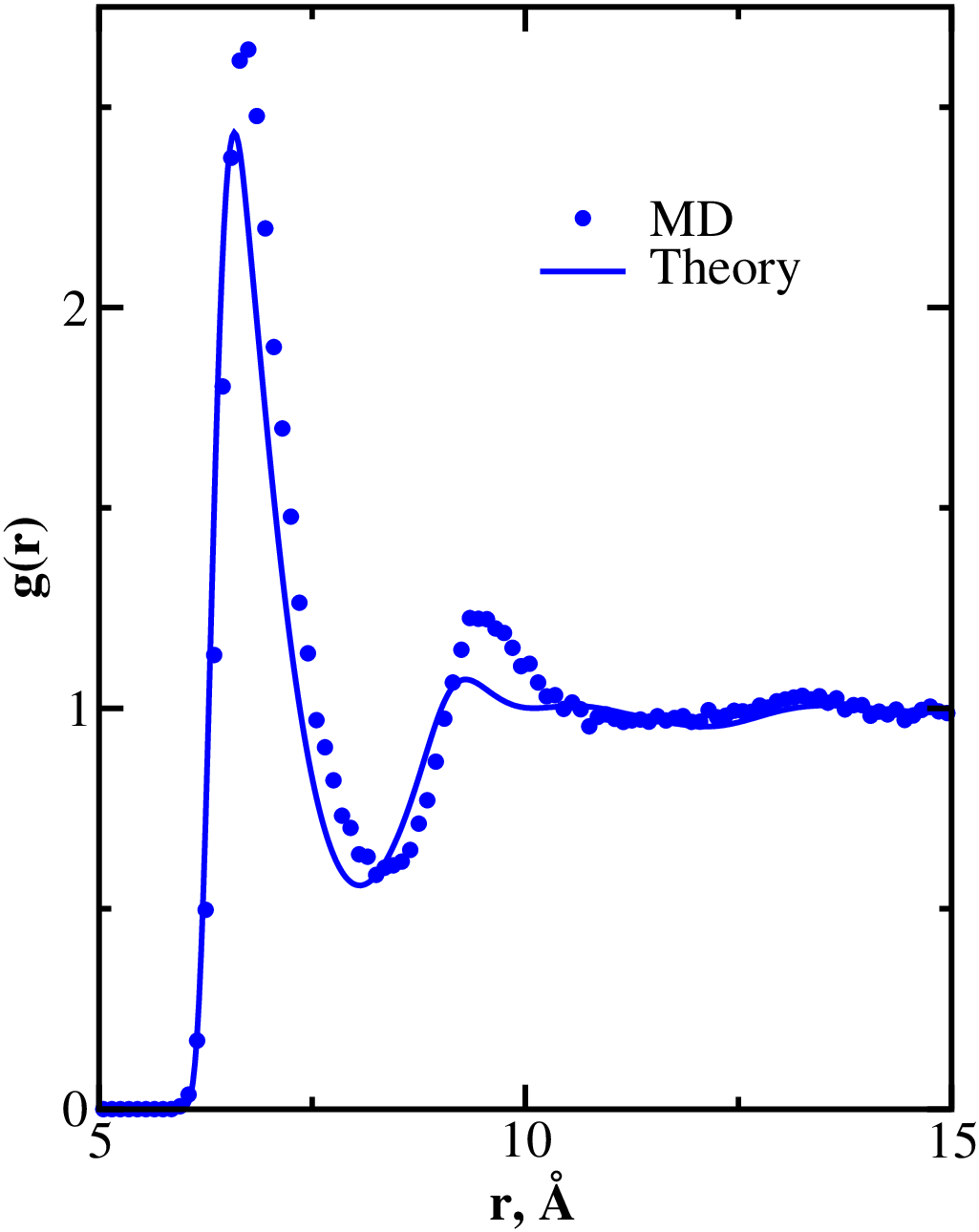}
\vspace{0.5cm}
\caption{Fullerene-water radial distribution function at T=300~K and 
$\rho$=1 g/cm$^3$. Symbols are the simulation results
  for the fullerene-water radial distribution function with all-atom SPC/E 
water,\cite{athawale09} and the line is from HNC theory for
coarse-grained water described by the LJG potential.}
\label{fig6}
\end{figure}

In Fig.~6 we compare simulation results
  for the fullerene-water radial distribution function with all-atom
  SPC/E  water,\cite{athawale09} and the HNC result for $g(r)$ 
with coarse-grained LJG water.
The agreement between the two distribution functions  is satisfactory
in general, although theory underestimates the heights of both
primary and secondary peaks of $g(r)$. Somewhat surprisingly, these
discrepancies do not manifest themselves in the results for the
solvent-mediated potential of mean force between two dilute fullerenes,
as can be seen from Fig.~7, where we present simulation\cite{li05b}
and theoretical results for $\Delta W(R)$. Theory and simulation are
in good agreement throughout the entire range of separations studied,
which could be a result of fortuitous cancellation of errors. 

\begin{figure}[htb]
\includegraphics[scale=0.5]{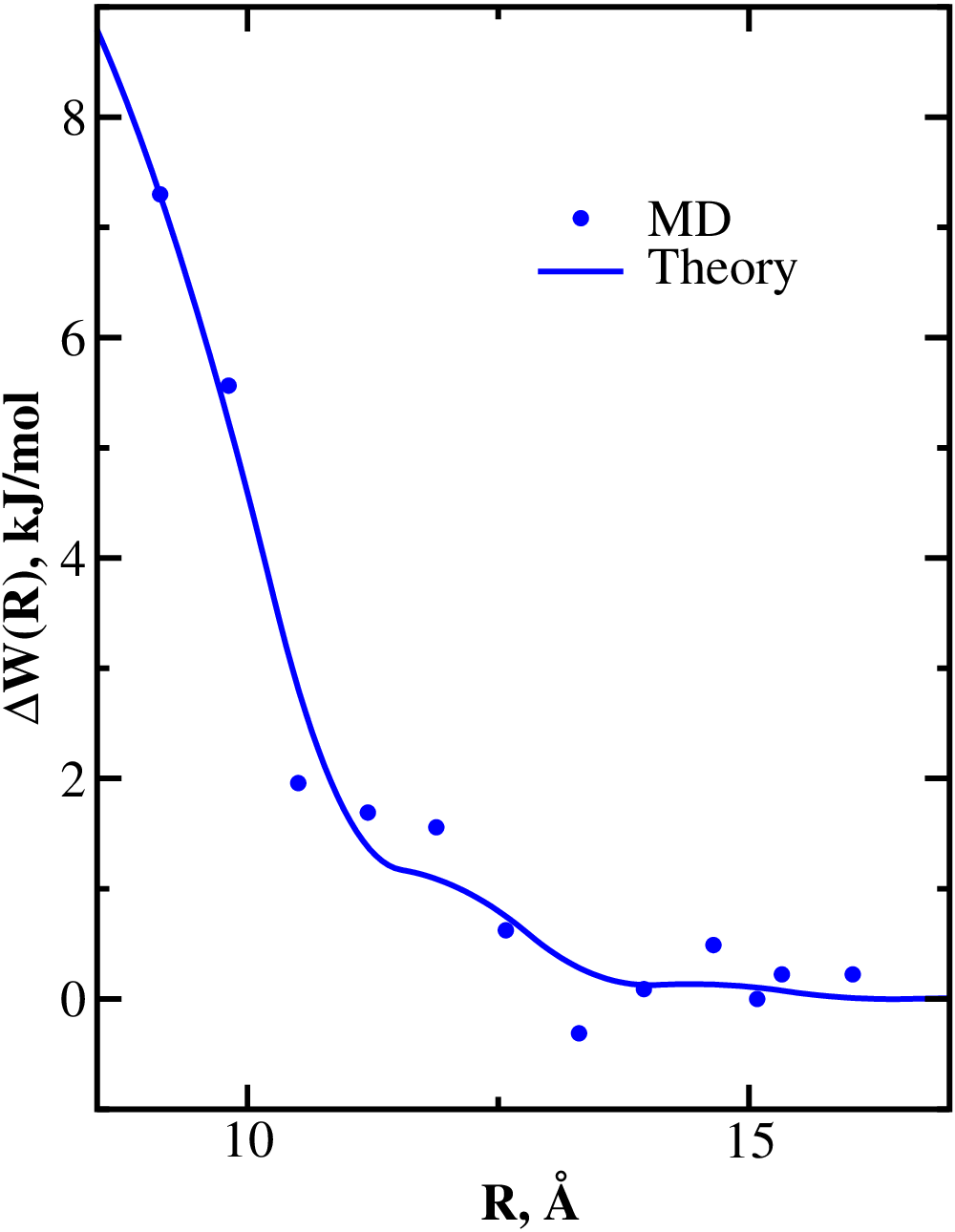}
\vspace{0.5cm}
\caption{Solvent-mediated PMF between two dilute fullerene solutes in
  water. Symbols are the simulation results,\cite{li05b} 
and lines are from integral equation theory.}
\label{fig7}
\end{figure}

In addition to the structural aspects of fullerene hydration, we have
also studied dynamics of fullerene solutes in water. In particular, we
have used the MCT approach outlined in the previous section to compute
the diffusion coefficient of fullerene in water at thermodynamic state
point T=300~K and $\rho$=1 g/cm$^3$; we have obtained the value 
$D$=1.73 nm$^2$/ns. This can be compared to the simulation 
result\cite{colherinhas11} of $D$=1.9 $\pm$ 0.4 nm$^2$/ns,
i.e. theoretical value is within the error-bars of the simulation
result. 

\section{Conclusion}
\label{sc4}

In this paper we have performed a theoretical study of structural,
thermodynamic, and dynamic properties of a coarse-grained water model
based on a spherically symmetric LJG pair potential.
For pure water, thermodynamically
self-consistent HMSA integral equation theory was shown to give 
accurate results for microstructure and thermodynamics, 
while MCT theory was in good agreement with the simulation data
for the self-diffusion coefficient, including its anomalous density
dependence. 

In studying the structure of dilute solutions of hydrophobic solutes, 
anisotropic HNC theory was used to calculate the solute-solute PMF and
the equilibrium constant for the conversion of the contact pair into the 
solvent-separated pair. Theory was shown to be in good agreement with
the available simulation data for both methane-methane and 
fullerene-fullerene PMF. It would be of interest to extend our
theoretical approach to larger solutes, such as graphene 
plates;\cite{howard08} this will be the subject of future research.

\section{Acknowledgment}
The author would like to thank Profs.~M.~S.~Shell and S.~Garde,  
and Drs.~A.~Chaimovich and S.~N.~Jamadagni 
for sending the simulation data. 

\end{document}